\documentclass[dvips]{acta}
\usepackage{supertabular,lscape,epsfig}
\usepackage{amssymb}
\usepackage{amsmath}

\usepackage{graphicx}
\usepackage[dvipsnames]{xcolor}

\usepackage{hyperref}
\usepackage{breakurl}

\usepackage{rotating}

\SetPages{0}{0}

\SetVol{0}{2021}

\begin{document}

\begin{Titlepage}
\Title{Ground-based observations of the ZZ~Ceti star HS~1625+1231}

\Author{C~s. K~a~l~u~p$^{1, 2, 3}$,\ \ Z~s. B~o~g~n~\'a~r$^{1, 3}$ and\ \ \'A. S~\'o~d~o~r$^{1, 3}$}
{$^1$Konkoly Observatory, Research Centre for Astronomy and Earth Sciences, ELKH, H-1121 Konkoly Thege Mikl\'os \'ut 15-17, Budapest, Hungary\\
$^2$ELTE E\"otv\"os Lor\'and University, Department of Astronomy, P\'azm\'any P\'eter s\'et\'any 1/A, Budapest 1117, Hungary\\
$^3$MTA CSFK Lend\"ulet Near-Field Cosmology Research Group\\

e-mail:kalup.csilla@csfk.org, bognar.zsofia@csfk.org, sodor.adam@csfk.org}

\Received{Month Day, Year}
\end{Titlepage}

\Abstract{We present the results of our detailed light curve analysis of the ZZ~Ceti star HS~1625+1231. We collected photometric time series data at Konkoly Observatory on 14 nights, and performed Fourier analysis of these data sets. We detected 11 significant frequencies, where six of them are found to be independent pulsation modes in the 514\,--\,881\,s period range. By utilising these frequencies, we performed preliminary asteroseismic investigations to give constraints on the main physical parameters, and to derive seismic distances for the star. Finally, we compared the astrometric distance provided by the \textit{Gaia} EDR3 data with those seismic distances. Our selected model, considering both the spectroscopic measurements and the distance value provided by \textit{Gaia}, has $T_{\mathrm{eff}} = 11\,000\,$K and $M_* = 0.60\,M_{\odot}$.}{techniques: photometric --
            stars: individual: HS~1625+1231 --
            stars: interiors --
            stars: oscillations -- 
            white dwarfs}

\section{Introduction}

White dwarfs (WDs) represent the final evolutionary stage of 97\% of the stars that populate the Universe. These earth-sized compact remnants are gradually cooling, therefore they have a wide range of temperature (200\,000\,K $\gtrsim T_{\mathrm{eff}} \gtrsim 4000$\,K). Their mass also spread a broad interval ($0.15 \lesssim M_{\mathrm{WD}}/M_{\odot} \lesssim 1.3$), but most of them are between $M_{\mathrm{WD}} \sim 0.5\,M_{\odot}$ and $M_{\mathrm{WD}} \sim 0.7\,M_{\odot}$. As a result of the high surface gravity ($\mathrm{log}~g \sim 8$), the so-called gravitational settling effect produces stratified structure, where heavier components sink, and lighter elements are on the surface. Based on the dominant element of the surface, we can distinguish several spectral types. The most populous group is that of DA type, which consists the $80\%$ of the white dwarfs, and their atmosphere has a thin hydrogen layer above the helium one. Many of them go through a phase of pulsational instability, when they turn into low-amplitude, multi-periodic pulsators with light variations on a time scale of minutes caused by surface temperature changes. These so-called DAV or ZZ~Ceti stars occupy a well-defined region in the Hertzsprung--Russel diagram between 10\,500\,K and 13\,000\,K effective temperatures. Oscillations detected in such objects are non-radial $g$-modes with periods ranging from 100\,s up to 1500\,s, typically with millimagnitude amplitudes. Both the classical $\kappa-\gamma$ driving mechanism (Dolez \& Vauclair 1981, Winget et al. 1982) in combination with the so-called convective driving mechanism (Brickhill 1991, Goldreich \& Wu 1999) are responsible for the excitation of the pulsation modes. However, they show a slightly different pulsational behaviour at different parts of the instability strip (Hermes et al. 2017). Hotter objects near the blue edge of the instability strip have stable amplitudes \mbox{($\sim$ 1\,mmag)} and phases, and have shorter periods (100\,--\,300\,s). At a few hundred degrees cooler stage, the periods are still short, but the amplitudes increase and reach the highest values in the middle of the instability strip. As the stars cool further, they can show irregularly recurring outbursts, when the stellar flux can increase up to 15\% (Bell et al. 2017). Closer to the red edge, variables are more likely to show short-term (from days to weeks) amplitude and frequency variations. Possible explanations behind this are resonant mode couplings (e.g. in Zong et al.\ 2016), interactions of pulsation and convection (e.g. Montgomery et al.\ 2010), or merely insufficient frequency resolution of the data sets. Finally, as we arrive to the red edge of the instability strip, the amplitudes decrease, and we can find the longest pulsation period ZZ~Ceti stars. For comprehensive observational and theoretical reviews, see Winget \& Kepler\ (2008), Fontaine \& Brassard\ (2008), Althaus et al.\ (2010), Hermes et al.\ (2017), C\'orsico et al.\ (2019) and C\'orsico\ (2020).

Pulsating white dwarf stars likely have the same properties as their non-variable counterparts. Based on the fact that the pulsation modes are very sensitive to the global structure of the star, asteroseismic investigations are the only way we can study their internal structure. The more pulsational modes we detect, the stronger constraints we can provide through asteroseismology. Currently, we know about 260 ZZ~Ceti stars (C\'orsico 2020), but most of them have pulsation periods only from the sometimes short discovery runs. This makes the search for new pulsators and the follow-up observations of known white dwarfs an important effort: we need to detect more pulsation modes for asteroseismology to properly constraint the physical parameters of the stars, and better understand the properties of compact pulsators in general.

In this paper we present time-series photometry observations of a ZZ Ceti star HS 1625+1231. We carry out Fourier analyses to identify pulsational modes, and performed preliminary asteroseismical investigations to determine the physical parameters of the star. This publication is a part of our long-term ground-based observations on pulsating white dwarf stars. For previous studies see: Bogn\'ar et al.\ (2009, 2014, 2016, 2019, 2021), and Papar\'o et al.\ (2013).

\section{Observations and data reduction}

We collected photometric time-series data on the HS~1625+1231 variable ($G=16.27$\,mag, $\alpha_{2000}=16^{\mathrm h}28^{\mathrm m}13^{\mathrm s}$, $\delta_{2000}=+12^{\mathrm d}24^{\mathrm m}53^{\mathrm s}$) with the 1-m  Ritchey-Chr\'etien-Coud\'e telescope located at Piszk\'estet\H o Mountain Station of Konkoly Observatory, Hungary. The  measurements  were  made with an FLI Proline 16803 CCD camera in white light on 14 nights. We used exposure times between 40\,s and 60\,s depending on the weather conditions, while the read-out time of the camera was approximately 3\,s. Table~1 shows the journal of observations of HS~1625+1231, and Fig.~1 represents a plot of these ground-based data as normalised differential light curves.

\MakeTable{lrccrrr}{\textwidth}{Journal of observations of HS~1625+1231. `Exp' is the integration time used, \textit{N} is the number of data points, and $\delta T$ is the length of the data sets including gaps. Weekly observations are denoted by `a,b,c,d,e' letters in parentheses.}
{\hline
Run & UT Date & Start time & Exp. & \textit{N} & $\delta T$ \\
 &  & (BJD-2\,450\,000) & (s) &  & (h) \\
\hline
01(a) & 2019 Mar 12 & 8555.498 & 45 & 260 & 3.45 \\
02(b) & 2019 Apr 06 & 8580.437 & 45 & 93 & 1.27 \\
03(b) & 2019 Apr 07 & 8581.426 & 60 & 223 & 4.92 \\
04(c) & 2019 May 31 & 8635.351 & 60 & 279 & 5.12 \\
05(c) & 2019 Jun 01 & 8636.343 & 40 & 372 & 5.20 \\
06(c) & 2019 Jun 02 & 8637.342 & 40 & 441 & 5.37 \\
07(c) & 2019 Jun 04 & 8639.351 & 40 & 414 & 5.06 \\
08(d) & 2019 Jun 27 & 8662.347 & 60 & 245 & 4.36 \\
09(d) & 2019 Jun 29 & 8664.348 & 60 & 241 & 4.56 \\
10(d) & 2019 Jun 30 & 8665.352 & 60 & 264 & 4.63 \\
11(e) & 2019 Jul 18 & 8683.327 & 40 & 320 & 3.82 \\
12(e) & 2019 Jul 22 & 8687.341 & 60 & 129 & 2.24 \\
13(e) & 2019 Jul 23 & 8688.326 & 60 & 180 & 3.15 \\
14(e) & 2019 Jul 24 & 8689.330 & 60 & 189 & 3.31 \\
\multicolumn{2}{l}{Total:} & & \multicolumn{2}{r}{3650} & 56.46\\
\hline
}

\begin{figure}
\centering
\includegraphics[width=1.0\textwidth]{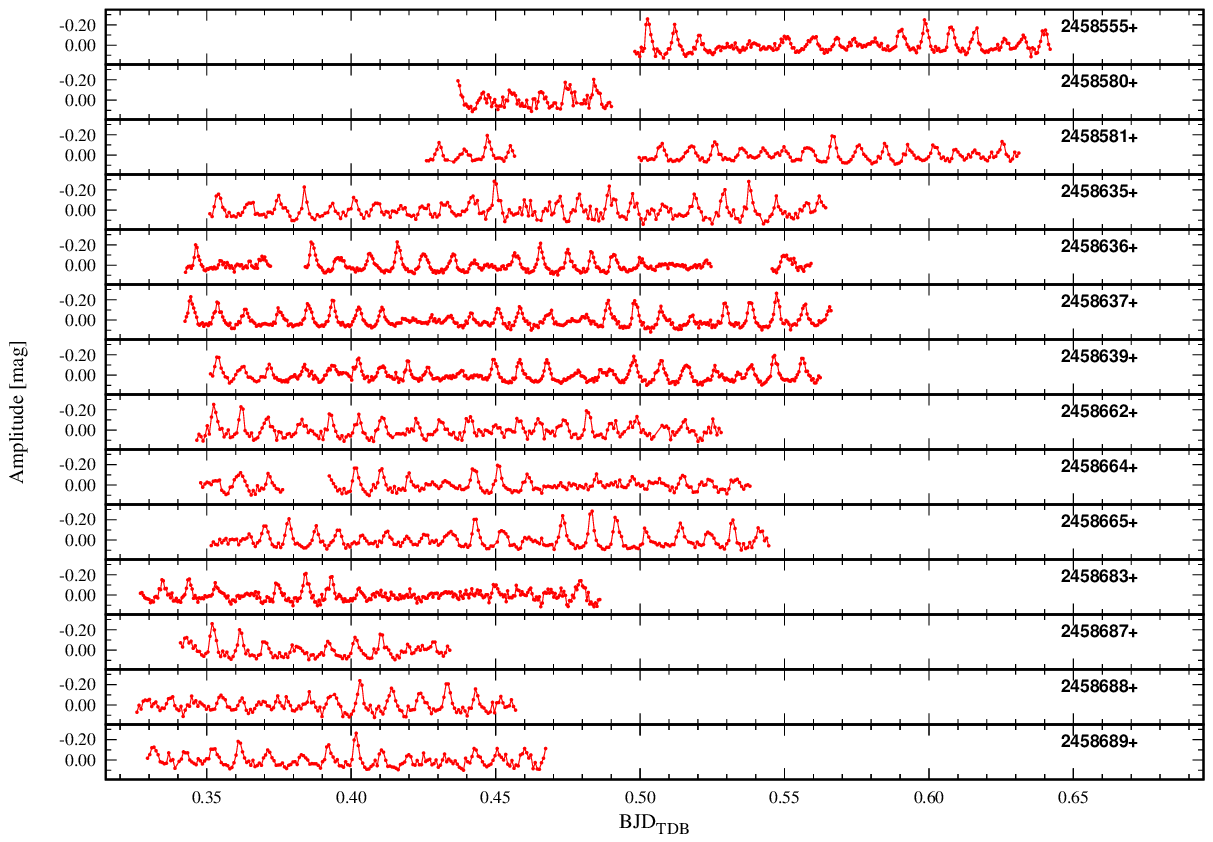}
\FigCap{Normalized differential light curves of HS~1625+1231.}
\end{figure}

To obtain the light curves ready for analysis, we reduced the raw data frames with \textsc{iraf}\footnote{\textsc{iraf} is distributed by the National Optical Astronomy Observatories, which are operated by the Association of Universities for Research in Astronomy, Inc., under cooperative agreement with the National Science Foundation.} tasks. We utilised them to perform standard bias, dark and flat corrections, then we employed aperture photometry of the target and some field objects. We used the average magnitudes of these latter, non-variable objects (2MASS 16282312+1222067 and 2MASS 16281789+1222006) as comparison stars for the differential photometry procedure. Then we corrected for the low-frequency atmospheric and instrumental effects by fitting a second- or third-order polynomial to the resulting differential light curves. This procedure did not affect the known frequency domain of pulsating ZZ~Ceti stars. We also converted the observation times of every data point to barycentric Julian dates in barycentric dynamical time ($\mathrm{BJD_{TDB}}$) using the applet of Eastman et al.\ (2010)\footnote{http://astroutils.astronomy.ohio-state.edu/time/utc2bjd.html}.

\section{Light curve analysis}

We performed Fourier analysis of the daily, weekly, and the whole data sets using the photometry modules of the Frequency Analysis and Mode Identification for Asteroseismology (\textsc{famias}) software package (Zima 2008). We used the standard consecutive prewhitening technique to determine the possible pulsational frequencies by the Fourier periodograms of the data set. We also constructed various subsets from different consecutive weekly observations, to test our pulsational frequency solutions and to reveal if there is any $\pm$1\,d$^{-1}$ ambiguity due to the single-site ground-based observations. These subsets consisted of the data of weeks (a+b+c), (b+c+d), (c+d+e), (a+b+c+d), (b+c+d+e), each of them have different spectral windows. In every case, we accepted a peak significant if its amplitude exceeded a signal-to-noise ratio (S/N) of 5, following the method presented in Bogn\'ar et al.\ (2021). In practice, we found that as we approach the 4~S/N, we detect more and more closely spaced peaks with similar amplitudes. Thus, we decided to select the highest amplitude peaks which has higher probability to be intrinsic pulsation frequencies. We chose this conservative approach to avoid false positives.

The accepted frequencies are listed in Table~2, together with the frequencies derived from the whole data set. The noise level was calculated by the average Fourier amplitude in a $\sim 1700\,\mu$Hz radius vicinity ($150\,$d$^{-1}$) of the peak in question.

We note that we can detect additional frequencies with lower amplitudes around 1135, 1198, 1348, and 1434\,$\mu$Hz. These may originate from short-term amplitude/phase/frequency variations. Another possible interpretation is that some of them are rotational split frequency components. These components are situated in the 700\,--\,800-sec period range, where the modes are not stable enough in amplitude/phase/frequency, and prewhitening can leave residuals above the detection threshold, see e.g. Hermes et al.\ (2017) or Bogn\'ar et al.\ (2020). This phenomenon is not uncommon in pulsating white dwarfs, and sometimes the authors use Lorentzian fit to a broadened frequency group to determine a single underlying pulsation period. According to the Rayleigh frequency resolution, these closely spaced frequencies are resolved, but considering the findings described above, we cannot consider each and every significant signal as an independent pulsation mode. There is certainly a chance that a couple of real pulsation frequencies remain disregarded this way, but we chose this conservative approach to accept only reliable components for the asteroseismic analysis.

For benchmark reasons, we also performed the frequency analysis of the combined weekly data sets and the whole data set with \textsc{Period04} (Lenz \& Breger 2005) and \textsc{LCfit} (S\'odor 2012), but we did not find any difference between the results, and the closely spaced peaks were also detected at the same locations.

Based on the analysis of the whole data set, we found 11 frequencies in the \mbox{$\sim 230-3800\,\mu$Hz} range listed in Table~3. We identified 5 combination peaks among them, therefore we consider the remaining 6 peaks as independent pulsation modes suitable for asteroseismic investigations. As Table~2 shows, there are no aliasing ambiguities in the base frequencies among the subsets. We note however, that in the single case of $f_2+f_3$, the highest-ampitude peak is actually situated at $f_2+f_3-1$\,d$^{-1}$, which, according to the previous findings, is clearly a daily alias of $f_2+f_3$, which we accepted as the real intrinsic combination frequency.
Summarising our findings on the frequencies $f_1 - f_6$, based on the analysis of the whole data set:

\begin{itemize}
    \item $f_1$: the dominant frequency, with closely spaced peaks at $1198.48$ and $1216.0\,\mu$Hz.
    \item $f_2$: the second largest amplitude frequency, with closely spaced peak at $1422.79\,\mu$Hz.
    \item $f_3$: the third largest amplitude frequency, closely spaced pair at $1134.06\,\mu$Hz.
    \item $f_4$: a stable frequency as we expected according to the findings of Hermes et al.\ (2017), no closely spaced peaks.
    \item $f_5$: the fifth largest amplitude frequency wit closely spaced peaks at $1357.71$ and $1344.04\,\mu$Hz.
    \item $f_6$: this peak was found in two data sets only, but with a consistent frequency value, thus we accepted it as an independent pulsation mode. No closely spaced peaks. 
\end{itemize}

\MakeTable{lrrrrrrr}{\textwidth}{The accepted pulsation frequencies of HS~1625+1231 in different combined weekly data subsets.}
{\hline
 & \multicolumn{7}{c}{Frequency [$\mu$Hz]} \\
\hline
week(a+b+c)	&	& 1134.66 & 	& 1198.11	& 1347.80	& 1434.69	& 1946.77\\
week(b+c+d)	&	& 1134.87 &	    & 1198.13	& 1346.57	& 1434.70	& 1946.77\\
week(c+d+e)	&	236.56	& 1135.24	&	& 1198.18	& 1346.61	& 1434.73	& 1946.78\\
week(a+b+c+d)	&	&	1134.86	& & 1198.12	& 1347.38	& 1434.70	& 1946.77\\
week(b+c+d+e)	& 236.54 & 1135.25	& 1179.73	& 1198.15	& 1346.16	& 1434.70	& 1946.77\\
whole data set	&	236.56 &	1135.25	& 1179.73	& 1198.14	& 1348.64	& 1434.70 &	1946.77\\
\\
week(a+b+c)	& &	& 2632.94 & \\
week(b+c+d)	& 2332.57	& & 2632.92 & \\
week(c+d+e)	& 2332.95	& 2569.98	& 2632.90	& 3768.15 \\
week(a+b+c+d)	&	2332.57	&	& 2632.94 & \\
week(b+c+d+e)	&  2332.96	& 2569.51	& 2632.91	& \\
whole data set	&	2332.97	& 2558.35	& 2632.93	& 3768.16 \\
\hline
}

\MakeTable{lrrr}{\textwidth}{Frequencies, periods and amplitudes of the six independent pulsation components and their combination peaks based on the Fourier analysis of the whole data set.}
{\hline
 & \multicolumn{1}{c}{$f$} & \multicolumn{1}{c}{$P$} & \multicolumn{1}{c}{$A$} \\
 & \multicolumn{1}{c}{[$\mu$Hz]} & \multicolumn{1}{c}{[s]} & \multicolumn{1}{c}{[mmag]} \\
\hline
$f_1$	& 1198.14	& 834.63	& 41.3 \\
$f_2$	& 1434.70	& 697.01	& 33.2 \\
$f_3$	& 1135.25	& 880.86	& 23.9 \\
$f_4$	& 1946.77	& 513.67	& 17.1 \\
$f_5$	& 1348.64	& 741.49	& 16.7 \\
$f_6$	& 1179.73	& 847.65	& 6.6 \\\\
$f_1+f_2$	& 2632.93	& 379.81	& 12.1 \\
$f_1+f_3$	& 2332.97	& 428.64	& 9.9 \\
$f_2+f_3$	& 2569.50	& 389.18	& 7.7 \\
$f_2-f_1$	& 236.56	& 4227.23	& 7.8 \\
$f_1+f_2+f_3$	& 3768.16	& 265.38	& 5.7 \\
\hline
}

The prewhitening process is detailed in Fig.~2. Figure~3 shows the Fourier periodogram of the whole data set with its prewhitened periodogram.

\begin{figure}
\centering
\includegraphics[width=1.0\textwidth]{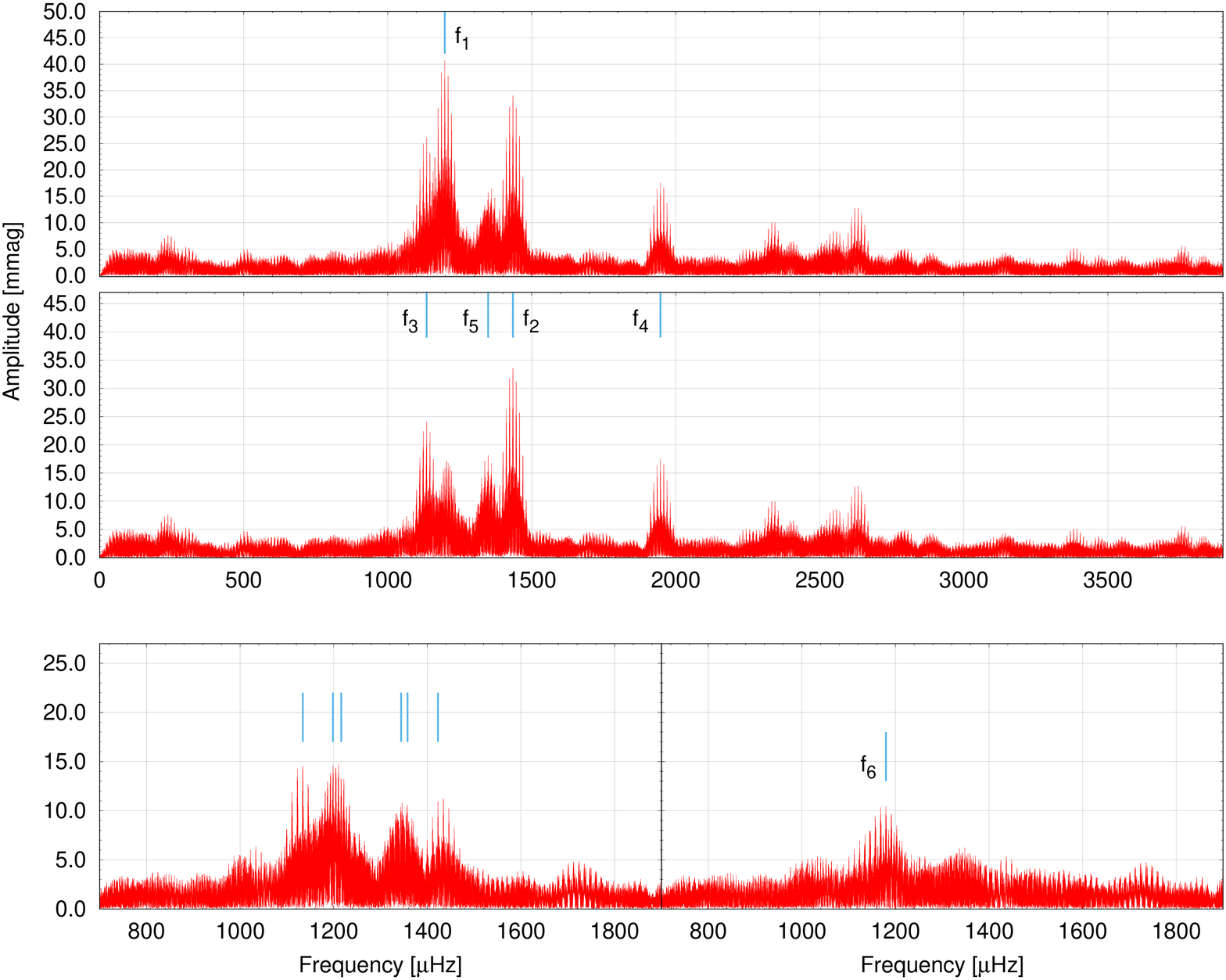}
\FigCap{Prewhitening process of the whole data set. The top panel shows the Fourier periodogram of the whole data set. The second panel reveals the periodogram after prewhitening with $f_1$. In the bottom left-hand panel we mark the five closely spaced peaks reviewed in the text, while the right-hand panel shows the periodogram after prewhitening with these frequencies.}
\end{figure}

\begin{figure}
\centering
\includegraphics[width=1.0\textwidth]{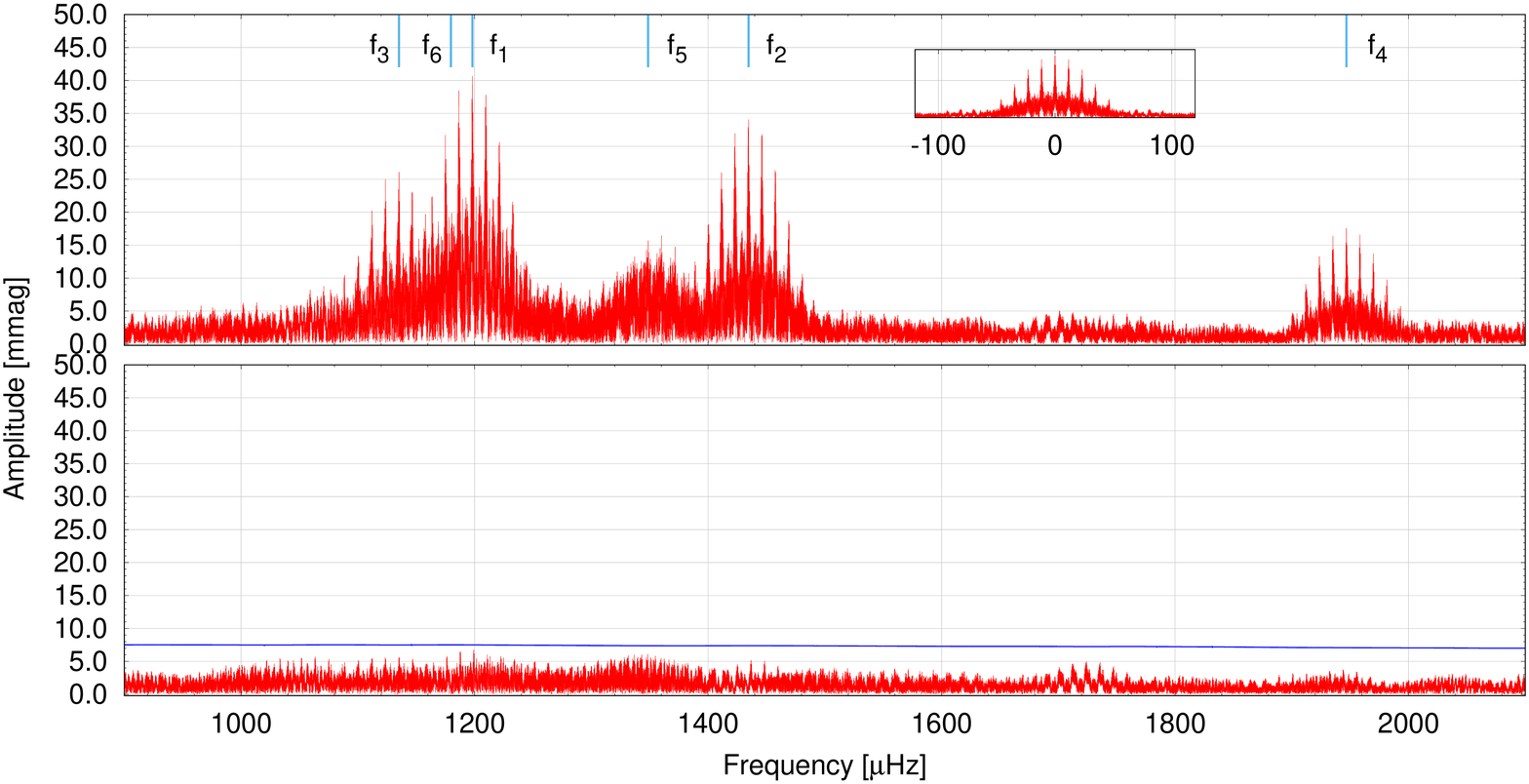}
\FigCap{The Fourier periodogram of the whole data set (top panel), with its prewhitened periodogram (bottom panel). We mark the accepted frequencies listed in Table~3 with blue lines. We denoted the S/N=5 significance level with a blue line in the bottom panel. The window function is shown in the inset.}
\end{figure}

\section{Asteroseismology}

To perform the asteroseismic analysis of the star, we utilised the same software and followed the same steps as we did in the case of the two ZZ~Ceti stars PM~J22299+3024 and LP~119-10 (Bogn\'ar et al. 2021). Our model grids were built by the latest (2018) version of the White Dwarf Evolution Code (\textsc{wdec}, Bischoff-Kim \& Montgomery 2018), which uses Modules for Experiments In Stellar Astrophysics (\textsc{mesa}, Paxton et al. 2011, 2013, 2015, 2018, 2019, version r8118) equation of state and opacity routines. We take a hot ($\sim100\,000$\,K) polytrope as a starting model, which is then evolved down to the temperature we request. The finally obtained model is a thermally relaxed solution to the stellar structure equations. We treat the convection according to the mixing length theory (Bohm \& Cassinelli 1971), and chose to use the $\alpha$ parametrisation, considering the results of Tremblay et al.\ (2015). 

Following the adiabatic equations of non-radial stellar oscillations (Unno et al. 1989), we computed the set of possible $\ell=1$ and $2$ eigenmodes for each model. Utilising the \textsc{fitper} program of Kim\ (2007), we then calculated the goodness of the fit between the observed ($P_i^{\mathrm{obs}}$) and calculated ($P_i^{\mathrm{calc}}$) periods. The quality of the fits is characterised by the root mean square ($\sigma_\mathrm{{rms}}$) value calculated for every model with the equation as follows:

\begin{equation}
\sigma_\mathrm{{rms}} = \sqrt{\frac{\sum_{i=1}^{N} (P_i^{\mathrm{calc}} - P_i^{\mathrm{obs}})^2}{N}}
\label{equ1}
,\end{equation}

\noindent where \textit{N} is the number of observed periods.

In a star rich in pulsation modes, one could try and identify the $\ell$ values of the modes before the model fits, relying on the approximately equidistant period spacings. Similarly, detecting rotationally split frequencies (triplets for $\ell=1$ and quadruplets for $\ell=2$ modes) could also help the preliminary identification of the modes. However, in our case, the number of only six independent modes is too low to find regular period spacings, and we also did not find signs of rotationally split frequencies. Therefore, we have to rely on the modelling itself for mode identification, which is common practice for such frequency-poor white dwarf pulsators, although undeniably adds more ambiguity to the results.

\subsection{Period fits}

At first, we performed the period fits utilising a coarse (master) grid, which covers a wide parameter space in effective temperature and stellar mass. The physical parameters we varied building this grid: $T_{\mathrm{eff}}$, $M_*$, $M_\mathrm{{env}}$ (the mass of the envelope, determined by the location of the base of the mixed helium and carbon layer), $M_\mathrm{H}$, $X_\mathrm{{He}}$ (the helium abundance in the C/He/H region), and $X_\mathrm{O}$ (the central oxygen abundance). Table~4 lists the parameter space we covered by the master grid and the corresponding step sizes.

$M_{\mathrm{He}}$ is the mass of the helium layer. We fixed its value at $10^{-2}\,M_*$, which is the theoretical maximum for this parameter, because of two reasons: the first is to reduce the number of free parameters and speed up the building of the model grids; the second is that considering the results of Romero et al.\ (2012), the mass of the He layer can be as much as a factor of 3$-$4 lower than the values according to evolutionary calculations, but not orders of magnitudes lower, which would affect the periods substantially.

The best-fit model was found to be at $T_{\mathrm{eff}} = 12\,500\,$K and $M_*=0.70\,M_{\odot}$ ($\sigma_\mathrm{{rms}} = 0.90\,$s). In this case, four out of the six modes was found to be $\ell=1$, however, the dominant frequency is $\ell = 2$. As a next step, we investigated, what solutions we obtain if we assume that half of the modes are $\ell=1$, including the dominant mode, taking into account the better visibility of $\ell = 1$ modes over $\ell = 2$ ones (see e.g. Castanheira \& Kepler 2008 and references therein).
In this case, the best-fit model has $T_{\mathrm{eff}} = 11\,000\,$K and $M_*=0.60\,M_{\odot}$, with $\sigma_\mathrm{{rms}} = 0.99\,$s, see Fig.~4. This means a $1500\,$K cooler and a less massive solution than we obtain if the dominant mode is $\ell=2$.

\begin{figure}
\centering
\includegraphics[width=1.0\textwidth]{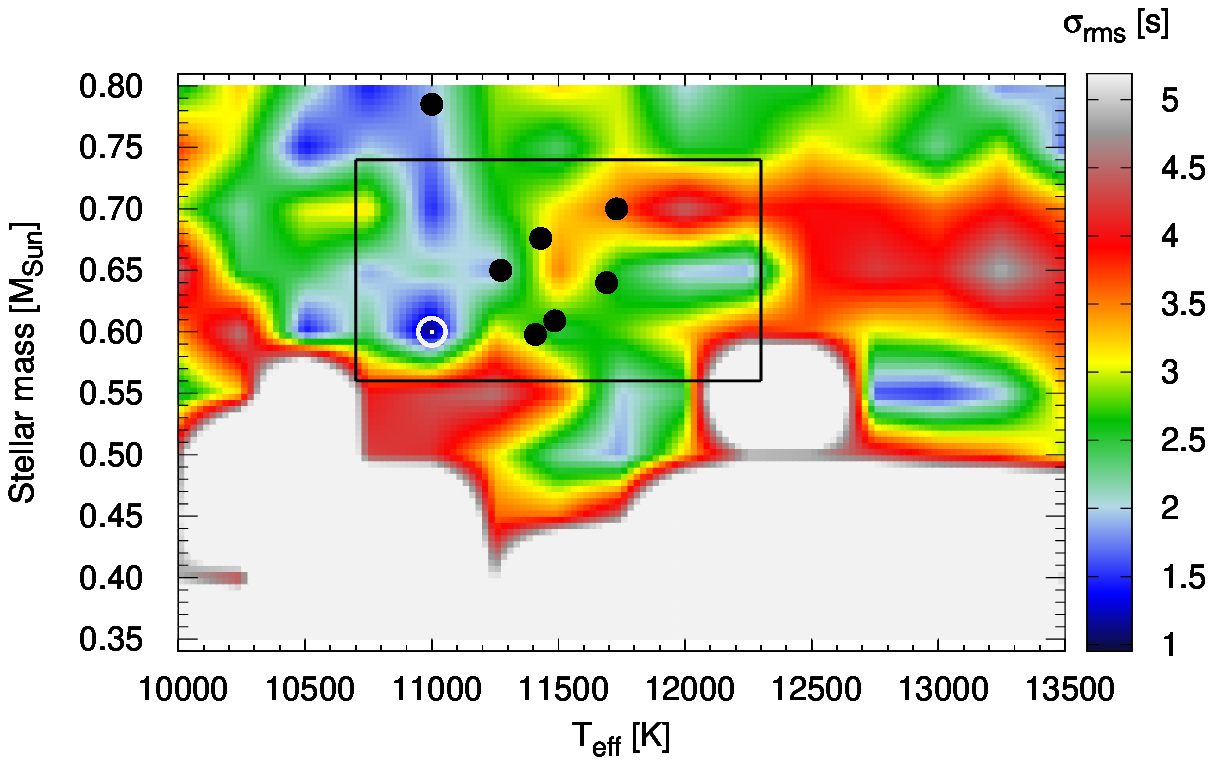}
\FigCap{Models on the $T_{\mathrm{eff}} - M_*$ plane utilising the master grid and assuming that half of the modes are $\ell=1$, including the dominant mode. The model with the lowest $\sigma_\mathrm{{rms}}$ value is denoted with a white open circle, while the spectroscopic solutions are signed with black dots. We also marked the parameter space further investigated by the refined grid with a black square.}
\end{figure}

Considering that it is more likely that the dominant mode is $\ell=1$, we performed different period fits under this assumption and utilising a refined grid in effective temperature, stellar mass, and mass of the hydrogen layer, covering the parameter space listed in Table~4.
If the dominant mode would be an $\ell=2$, it would imply a much larger physical amplitude for this mode since it has the largest light amplitude. That is, it is more likely that the dominant pulsation is coming from an $\ell=1$ mode.

Assuming that the dominant mode is $\ell=1$ and there are three $\ell=1$ modes, the best-fit model has $T_{\mathrm{eff}} = 11\,000\,$K and $M_*=0.67\,M_{\odot}$ ($\sigma_\mathrm{{rms}} = 0.82\,$s). Imposing even more constraint by assuming that four $\ell=1$ modes exist, the best fit model is $T_{\mathrm{eff}} = 11\,000\,$K and $M_*=0.60\,M_{\odot}$ ($\sigma_\mathrm{{rms}} = 0.99\,$s). For curiosity, we performed another period fit not giving any restrictions on the $\ell$ values of the modes except that the dominant mode is $\ell=1$. In this case, the best-fit model parameters are $T_{\mathrm{eff}} = 11\,200\,$K and $M_*=0.74\,M_{\odot}$ ($\sigma_\mathrm{{rms}} = 0.55\,$s).  

\MakeTable{lrrr}{\textwidth}{Parameter spaces covered by the master grid and the refined grids. The step sizes are in parentheses.}
{\hline
& \multicolumn{1}{c}{Master grid} & \multicolumn{1}{c}{Refined grid} \\
\hline
$T_{\mathrm{eff}}$ [K] & $10\,000 - 13\,500$ [250] & $10\,700 - 12\,300$ [100]\\
$M_*$ [$M_{\odot}$] & $0.35 - 0.80$ [0.5] & $0.56 - 0.74$ [0.1]\\
-log$(M_\mathrm{{env}}/M_*)$ & $1.5 - 1.9$ [0.1] & $1.5 - 1.9$ [0.1]\\
-log$(M_{\mathrm{He}}/M_*)$ & $2$ [fixed] & $2$ [fixed]\\
-log$(M_\mathrm{H}/M_*)$ & $4 - 9$ [$1.0$] & $4 - 9$ [0.5]\\ 
$X_\mathrm{{He}}$ & $0.5 - 0.9$ [0.1] & $0.5 - 0.9$ [0.1]\\
$X_\mathrm{O}$ & $0.5 - 0.9$ [0.1] & $0.5 - 0.9$ [0.1]\\
\hline
}

Table~5 summarises the physical parameters of these best-fit solutions. It is conspicuous that $-$log$M_\mathrm{{env}} = 1.5$ in every case, and it would worth to try extending the grid to include lower values of $-$log$M_\mathrm{{env}}$. However, below this value the helium starts mixing with the core, and the equation of state in the core does not provide for anything other than carbon and oxygen. That is, venturing significantly below $-$log$M_{\mathrm{env}} = 1.5$ for $M_{\mathrm{env}}$, the tail of the helium in the C/He transition zone would bleed into the core (Agnes Bischoff-Kim, private communication).

\MakeTable{lccccccc}{\textwidth}{Physical parameters of the best-fit models.}
{\hline
$T_{\mathrm{eff}}$ [K] & $M_*$ [$M_{\odot}$] & -log$M_\mathrm{{env}}$ & -log$M_\mathrm{He}$ & -log$M_\mathrm{H}$ & $X_\mathrm{{He}}$ & $X_\mathrm{O}$ & $\sigma_\mathrm{{rms}}$ (s) \\
\hline
\multicolumn{8}{l}{master grid, $4\ell=1$, but the dominant mode is $\ell=2$} \\
12\,500 & 0.70 & 1.5 & 2.0 & 6.0 & 0.8 & 0.9 & 0.90 \\
\\
\multicolumn{8}{l}{master grid, $3$ or $4\ell=1$, including the dominant mode} \\
11\,000 & 0.60 & 1.5 & 2.0 & 4.0 & 0.6 & 0.5 & 0.99 \\
\\
\multicolumn{8}{l}{refined grid, $3\ell=1$, including the dominant mode} \\
11\,000 & 0.67 & 1.5 & 2.0 & 5.5 & 0.6 & 0.9 & 0.82 \\
\\
\multicolumn{8}{l}{refined grid, $4\ell=1$, including the dominant mode} \\
11\,000 & 0.60 & 1.5 & 2.0 & 4.0 & 0.6 & 0.5 & 0.99 \\
\\
\multicolumn{8}{l}{refined grid, no restrictions on the number of $\ell=1$ modes, except that the dominant is $\ell=1$} \\
\multicolumn{8}{l}{(two $\ell=1$ solutions)} \\
11\,200 & 0.74 & 1.5 & 2.0 & 8.0 & 0.8 & 0.9 & 0.55 \\
\hline
}

In sum, we can say that the effective temperature of the star could be around $11\,000 - 11\,200\,$K, but we cannot obtain such strong constraint on the mass of HS~1625+1231, as according to the period fits utilising the refined grid, we find acceptable solutions between $0.60$ and $0.74\,M_{\odot}$, too.

However, there is a way to validate our seismic findings: by comparing the astrometric distance provided by the \textit{Gaia} space mission (Gaia Collaboration 2016) for this star with the seismic distances calculated by the selected models. The steps on how to derive the seismic distances utilising the luminosity of the selected models and the apparent visual magnitude of the star are specified e.g. in Bell et al.\ (2019) or Bogn\'ar et al.\ (2021). Table~6 summarises the seismic distances derived for the different model solutions, together with the geometric distance value published by Bailer-Jones et al.\ (2021) based on the \textit{Gaia} early third release (EDR3; Gaia Collaboration 2021). We utilised the apparent visual magnitude of HS~1625+1231 published in the fourth US Naval Observatory CCD Astrograph Catalog (Zacharias et al. 2012): $m_{\mathrm V} = 16.154\pm0.08\,$mag.


\MakeTable{lccc}{\textwidth}{Seismic distances calculated for the different best-fit models utilising the master and refined grids, see Table~5.}
{\hline
$T_{\mathrm{eff}}$ [K] & $M_*$ [$M_{\odot}$] & $\mathrm{log}\,L/L_{\odot}$ & $d_{\mathrm{seismic}}$ [pc]\\
\hline
12\,500 & 0.70 & $-2.555$ & $72.7\pm2.7$ \\
11\,000 & 0.67 & $-2.752$ & $65.9\pm2.4$ \\
11\,000 & 0.60 & $-2.663$ & $73.0\pm2.7$ \\
11\,200 & 0.74 & $-2.796$ & $61.6\pm2.3$\\
        &      &          & $d_{Gaia} = 77.735^{+0.274}_{-0.288}$\\
\hline
}

Considering the effective temperature and surface gravity values in the literature for HS~1625+1231:
    \begin{enumerate}
        \item Voss et al.\ (2006): 11\,272\,K, 0.65\,$M_\odot$
        \item Fontaine \& Brassard\ (2008): 11\,730\,K, 0.7\,$M_\odot$
        \item Castanheira \& Kepler\ (2009): 11\,000\,K, 0.785\,$M_\odot$ -- modelling
        \item Gianninas et al.\ (2011); we corrected their $T_{\mathrm{eff}}$ and $\mathrm{log}\,g$ values according to the findings of Tremblay et al.\ (2013), based on radiation-hydrodynamics three-dimensional simulations of convective DA stellar atmospheres: 11\,690\,K, 0.64\,$M_\odot$
        \item Romero et al.\ (2012): 11\,485\,K, 0.609\,$M_\odot$ -- modelling
        \item Kepler et al.\ (2015): 11\,430\,K, 0.676\,$M_\odot$
        \item Gentile Fusillo et al.\ (2019): 11\,409\,K, 0.598\,$M_\odot$
    \end{enumerate}
    
    According to the literature values, the most probable effective temperature of the star is between 11\,000 and 11\,700\,K, however, for the stellar mass, we can find solutions in a relatively broad range from 0.6 to 0.79\,$M_\odot$. Voss et al. and Fontaine \& Brassard published only surface gravities, which we converted to stellar mass using the model calculations of Bradley (1996, ApJ, 468, 350).
    
    Considering the model solutions of Castanheira \& Kepler (2009) and Romero et al., we have to mention that they used many of the periods listed in Voss et al., despite the fact that the authors marked most of those frequencies as probable results of cloud interference or alias peaks.
    
    Choosing the best model solution is not an easy task in this case. We consider this asterosesmic investigation as a preliminary one. Taking into account the a \textit{Gaia} meaurements, both the 12\,500 and 11\,000\,K effective temperature model could be acceptable with $\sim 73$\,pc seismic distances. However, considering the literature temperature values, the 11\,000\,K, 0.6\,$M_\odot$ solution seems to be more realistic. Table~7 lists the calculated and observed period values of this model.
    
    \MakeTable{lccc}{\textwidth}{Calculated and observed periods of the selected model (11\,000\,K, 0.6\,$M_\odot$) for HS~1625+1231.}
    {\hline
    Model periods [s] ($\ell, k$) & Observed periods [s] \\
    \hline
    512.4 (2,17) & 513.7 \\
    696.2 (1,14) & 697.0 \\
    741.1 (1,15) & 741.5 \\
    835.3 (1,17) & 834.6 \\
    849.1 (2,30) & 847.7 \\
    881.9 (1,18) & 880.9 \\
    \hline
    }

The main goal of our preliminary asteroseismic modelling is to infer the main physical parameters, such as the effective temperature and stellar mass. We do not intend to investigate the chemical compositions in details. However, we have to mention that the 50 per cent oxygen abundance might be too low, considering the evolutionary calculations of Romero et al. (2012, MNRAS, 420, 1462). We searched their database for a model close in physical parameters to our selected model and found that the central oxygen abundance might rather be around 0.72 instead of our solution of 0.5.

We plotted the chemical composition profiles and the corresponding Brunt--V\"ais\"al\"a frequency profile for the selected model in Fig.~5.
    
    \begin{figure}
    \centering
    \includegraphics[width=\textwidth]{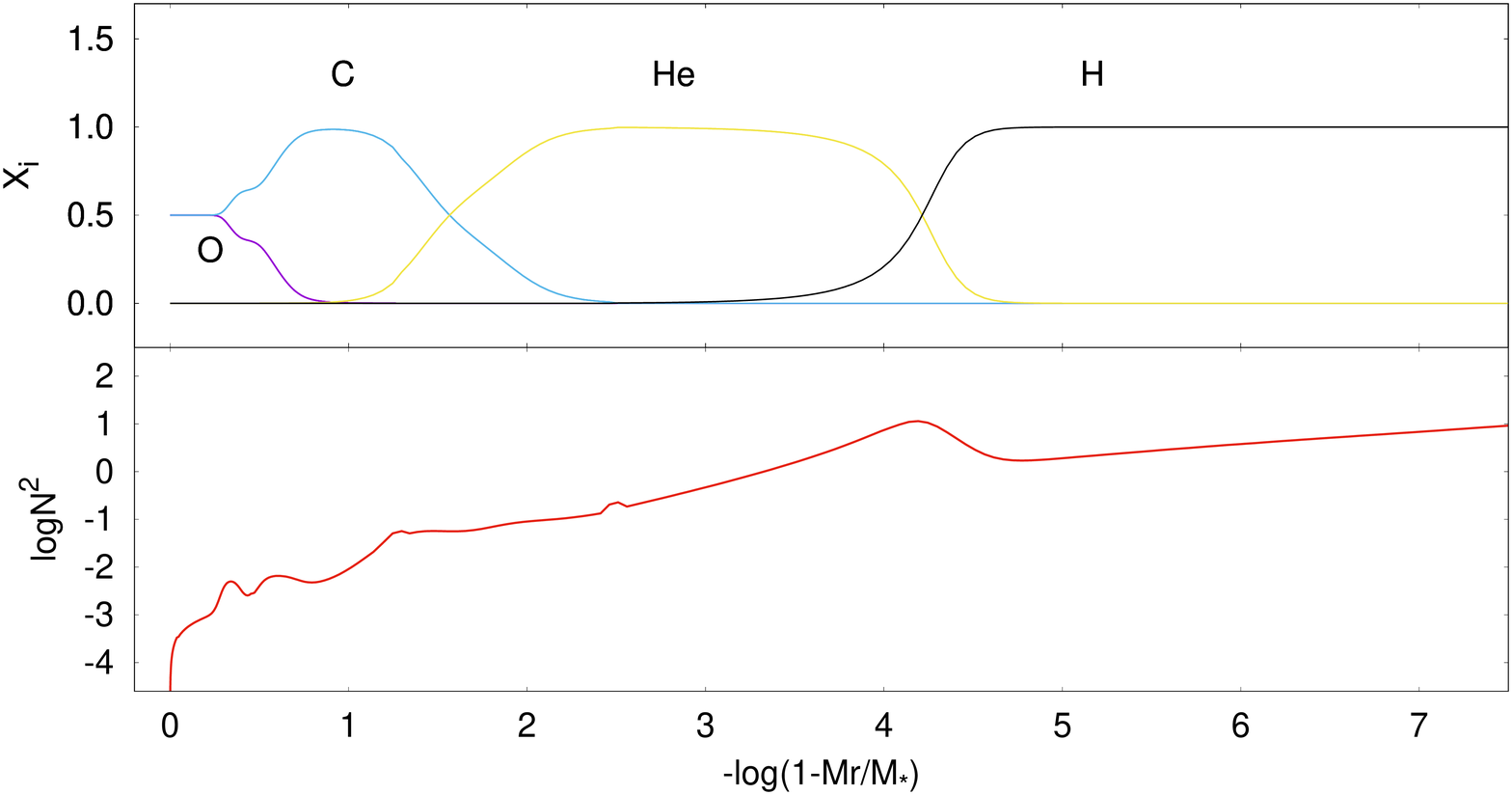}
    \caption{Chemical composition profiles (in fractional abundances) and the corresponding logarithm of the squared Brunt--V\"ais\"al\"a frequency ($\mathrm{log}\,N^2$) for the selected model ($T_{\mathrm{eff}}=11\,000\,$K, $M_*=0.60\,M_{\odot}$, $-$log$M_\mathrm{{env}}=1.5$, $-$log$M_\mathrm{He}=2.0$, $-$log$M_\mathrm{H}=4.0$, $X_\mathrm{{He}}=0.6$, $X_\mathrm{O}=0.5$.}
    \label{}
    \end{figure}

\section{Summary}

We presented ground based observations of the ZZ~Ceti star HS~1625+1231, which was reported as a new variable by Voss et al.\ (2006). They derived three pulsation frequencies by their observations at 385.2, 533.6, and 862.9\,s. These values are slightly different from our findings, but note that they utilised a short (5005\,s) observational run. Unfortunately, the Transiting Exoplanet Survey Satellite
(\textit{TESS}; Ricker et al. 2015) did not observe this target.

We obtained time-series photometry from 14 nights and performed Fourier analysis of all data sets to look for new pulsation periods. We found 6 independent pulsation modes in the 514\,--\,881 s period range, and performed asteroseismic investigations to infer the main physical parameters of the star via the latest version of the White Dwarf Evolution Code. Based on comparing the detected pulsation periods to stellar models, the effective temperature of the star is around 11\,000\,--\,11\,200\,K. For the mass of the star, we found acceptable solutions between 0.60 and 0.74\,$M_{\odot}$. We also calculated the seismic distances utilising the apparent visual magnitude of the star and the luminosity of the selected models. 
Our selected model, considering both the spectroscopic measurements and the distance value provided by \textit{Gaia}, has $T_{\mathrm{eff}} = 11\,000\,$K and $M_* = 0.60\,M_{\odot}$.


\Acknow{
The authors thank the anonymous referee for the constructive comments and recommendations on the manuscript.

The authors acknowledge the financial support of the Lend\"ulet Program of the Hungarian Academy
of Sciences, projects No. LP2018-7/2021 and LP2012-31, and the support by the KKP-137523 `SeismoLab' \'Elvonal grant of the Hungarian Research, Development and Innovation Office (NKFIH). Zs.B. acknowledges the support by the J\'anos Bolyai Research Scholarship of the Hungarian Academy of Sciences.

This work has made use of data from the European Space Agency (ESA) mission {\it Gaia} (\url{https://www.cosmos.esa.int/gaia}), processed by the {\it Gaia} Data Processing and Analysis Consortium (DPAC, \url{https://www.cosmos.esa.int/web/gaia/dpac/consortium}). Funding for the DPAC
has been provided by national institutions, in particular the institutions participating in the {\it Gaia} Multilateral Agreement.
}

\end{document}